\documentclass[10pt,twocolumn,reqno,amsmath,amssymb,aps,prb,superscriptaddress,floatfix]{revtex4-2}
\usepackage{graphicx} 
\usepackage{enumitem}
\usepackage{amsmath}
\usepackage{mhchem}
\usepackage{color}
\usepackage{xcolor}
\usepackage{siunitx}
\usepackage{hyperref}
\hypersetup{colorlinks=true,linkcolor=blue,citecolor=blue}
\usepackage[all]{hypcap}
\hypersetup{
    colorlinks=true,
    linkcolor=blue,
    citecolor=blue,
    urlcolor=blue
    }
\usepackage{placeins}
    
\begin{document}

\title{Polarization Controlled Supercurrent in Ferroelectric Josephson Junction}
\author{Yaozu Tang}
\affiliation{Kavli Institute of Nanoscience, Delft University of Technology, Lorentzweg 1, 2628 CJ Delft, The Netherlands}
\affiliation{Department of Quantum Nanoscience, Delft University of Technology, Lorentzweg 1, 2628 CJ Delft, The Netherlands}
\author{Mazhar N. Ali}
\affiliation{Kavli Institute of Nanoscience, Delft University of Technology, Lorentzweg 1, 2628 CJ Delft, The Netherlands}
\affiliation{Department of Quantum Nanoscience, Delft University of Technology, Lorentzweg 1, 2628 CJ Delft, The Netherlands}
\author{Gerrit E. W. Bauer}
\affiliation{WPI-AIMR \& IMR  \& CSIS, Tohoku University, 2-1-1 Katahira, Sendai 980-8577, Japan}
\affiliation{Kavli Institute for Theoretical Sciences, University of the Chinese Academy of Sciences, Beijing 10090, China}
\author{Yaroslav M. Blanter}
\affiliation{Kavli Institute of Nanoscience, Delft University of Technology, Lorentzweg 1, 2628 CJ Delft, The Netherlands}
\affiliation{Department of Quantum Nanoscience, Delft University of Technology, Lorentzweg 1, 2628 CJ Delft, The Netherlands}
\date{\today}


\begin{abstract}
Josephson junctions are essential devices in superconducting electronics and quantum computing hardware. Here we predict electrical control of the supercurrent in composite superconductor-insulator-ferroelectric-insulator-superconductor (S-I-FE-I-S) Josephson junctions. Inversion symmetry broken by unequal dielectric barrier thicknesses and/or  potentials converts ferroelectric polarization reversal into a substantial change of the critical current. With a WKB tunneling model we obtain non-volatile switching of the critical current with on-off efficiency up to 0.9 for physically realistic parameters. This can be achieved by optimizing the thicknesses and potential barriers of the insulating layers, as well as the thickness and dielectric constant of the ferroelectric layer. We also derive a compact linear expression for the critical current valid for small polarizations. Our results identify ferroelectric Josephson junctions as electrically programmable superconducting current switches for cryogenic memory and logic applications.
\end{abstract}

\maketitle

\section{Introduction}

The Josephson effect, where a supercurrent flows through a junction formed by two superconductors separated by a weak link \cite{josephson1962,josephson1965,tinkham2004}, is a cornerstone phenomenon in condensed matter physics. It underpins a wide range of applications, including superconducting qubits \cite{mooij1999,martinis2002,koch2007,makhlin2001,clarke2008}, SQUID sensors \cite{jaklevic1964,cohen1972,fagaly2006a}, and superconducting single-photon detectors \cite{peacock1996,chen2011,walsh2021}. The incorporation of diverse quantum materials (QMs), including semiconducting nanowires \cite{doh2005,vandam2006,mourik2012,chang2015}, ferromagnets \cite{ryazanov2001,kontos2002,sellier2004,bergeret2005,buzdin2005b,oboznov2006,buzdin2008,robinson2010,larkin2012,baek2014}, and topological materials \cite{rokhinson2012,pikulin2012,vanheck2012,houzet2013,schrade2015,flensberg2021}, has further expanded the functionality of Josephson junctions (JJs).

Ferroelectric materials, characterized by their spontaneous and switchable electric dipolar order, have been widely utilized in applications ranging from memory devices to sensors \cite{dawber2005,setter2006,scott2007,boscke2011a,boscke2011,martin2016,khan2020,mikolajick2021,schroeder2022}. However, their integration as weak links in Josephson junctions remains underexplored, and the interplay between ferroelectric polarization and superconductivity has not yet been fully understood. A notable feature of ferroelectric materials is the giant tunneling electroresistance (TER) observed in ferroelectric tunnel junctions (FTJs), in which a nanometer-thick ferroelectric layer acts as the tunneling barrier between metallic electrodes \cite{zhuravlev2005,garcia2009,maksymovych2009,gruverman2009}. This implies that ferroelectric Josephson junctions could exhibit unique properties with potential applications in cryogenic memory and computation \cite{paghi2025}.

Recently, directional supercurrent rectification-the superconducting diode effect-has attracted considerable attention \cite{nadeem2023,ando2020,wu2022,baumgartner2022,trahms2023}. This effect typically relies on broken inversion and time-reversal symmetries, often via magnetic textures or fields \cite{baumgartner2022,trahms2023,pal2022}. However, for scalable and power-saving device applications, local voltage control is highly desirable. Here, we propose an all-electrical route to strongly modulate (switch) the critical current via electric polarization, which offers an alternative for controllable superconducting devices without invoking magnetism. 

We analyze a composite S-I-FE-I-S junction in which a ferroelectric layer is sandwiched between two (para-/di-electric) insulating layers with independently tunable thicknesses and potential barrier heights. The reversal of a perpendicular electrical polarization shifts the electrostatic potential profiles. With broken inversion symmetry of the stacking, the normal state conductance and thus the critical current differ for different polarization directions. We find a polarization-controlled on-off efficiency that can approach 0.9 for physically realistic parameters and identify trade-offs with absolute current magnitude. We also derive a compact linear formula for the polarization dependence of the critical current valid at small polarization, offering rapid estimates for device modelling and design.

The manuscript is organized as follows: In Sec.~\ref{sec:model}, we present our tunneling model of the composite ferroelectric Josephson junction. In Sec.~\ref{sec:polarization control}, we discuss the polarization dependence of the critical current as well as how the on-off efficiency can be tuned by adjusting the system parameters. In Sec.~\ref{sec:analytical}, we present analytical expressions of the critical current in the limit of small polarization. Finally, in Sec.~\ref{sec:conclusion}, we summarize our findings and present an outlook.

\section{S-I-FE-I-S model}
\label{sec:model}

\begin{figure}[ht]
    \centering
    \includegraphics[width=0.9\columnwidth]{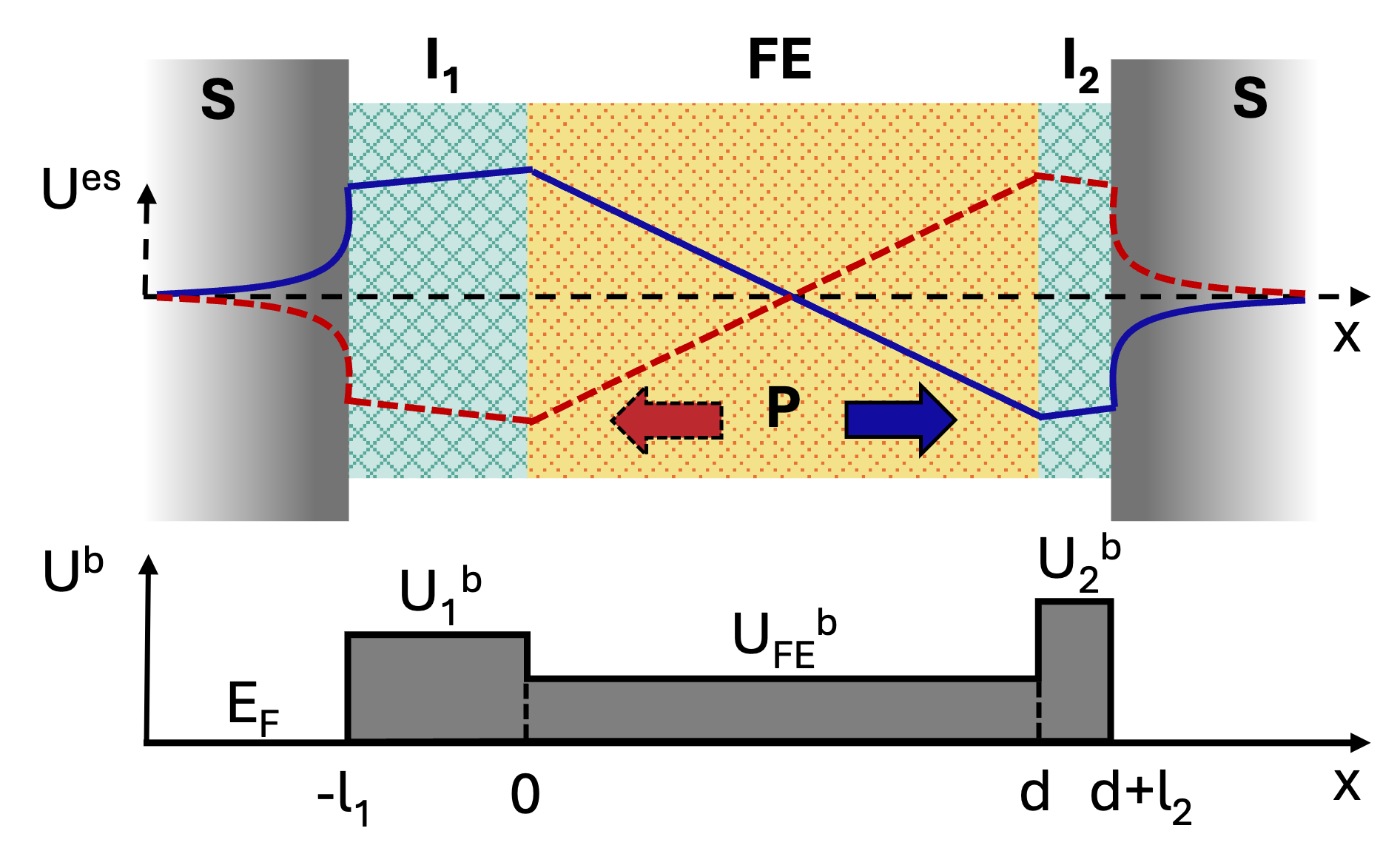}
    \caption{\textbf{S-I-FE-I-S composite ferroelectric junction.} Top: Geometry and electrostatic potential $U^{es}$ for opposite polarization directions (blue vs. red). Bottom: An asymmetric potential barrier profile $U^{b}$ in the dielectric state.}
    \label{fig:system}
\end{figure}

We consider a composite superconductor-insulator-ferroelectric-insulator-superconductor (S-I\(_1\)-FE-I\(_2\)-S) junction (Fig.~\ref{fig:system}) with broken inversion symmetry as a minimal model. The non-ferroelectric electrically insulating layers, $I_1$ and $I_2$, form potential barriers with heights $U^{b}_{1}$ and $U^{b}_{2}$ and thicknesses $l_1$ and $l_2$. We assume here that the superconductor (S) on both sides are the same. The ferroelectric (FE) is an electrical insulator with thickness d that forms a potential barrier $U^b_f$ in its dielectric state (see bottom panel of Fig.~\ref{fig:system}). Its polarization $P$  is uniform and normal to the interfaces.  The polarization induces equal and opposite bound charges at the FE interfaces, which are screened by free carriers in the metallic electrodes \cite{zhuravlev2005}. The resulting electrostatic potential $U^{es}$ is linear inside the ferroelectric and decays exponentially in the metals over the (Thomas-Fermi) screening length. The total potential profile entering the tunneling problem is the sum
\[
U(x) = U^{es}(x)+U^{b}(x).
\]
We derive in the following that when inversion symmetry is broken, reversing $P$ changes $U^{es}$, the electrical conductance and the critical current.

The electrostatic potential in the metallic electrodes is described by the Thomas-Fermi screening model \cite{mehta1973,zhuravlev2005}:
\begin{equation}
    U^{es}(x)=
    \begin{cases}
        \varepsilon_0^{-1}\sigma_s \delta\, e^{-|x+l_1|/\delta}, & x\le -l_1,\\[4pt]
        -\varepsilon_0^{-1}\sigma_s \delta\, e^{-|x-(d+l_2)|/\delta}, & x\ge d+l_2,
    \end{cases}
    \label{eq:Thomas_Fermi_screening}
\end{equation}
where $\sigma_s$ is the surface screening charge density and $\delta$ the screening length. For normal metals $\delta$ is the Thomas-Fermi length $\delta_{TF}^2=2\varepsilon_0 E_F/3e^2 n$, and is typically below 1 nanometer for high-density metals. Here $E_F$ is the Fermi energy and $n$ the electron density. In superconductors, the appropriate screening length is  being debated: Whether it remains the same as the normal-state Thomas-Fermi length or is given by the much larger London penetration depth \cite{hirsch2003,hirsch2004,hirsch2004a,desimoni2018}. In this paper we adopt $\delta=\delta_{TF}$ for the main calculations and comment on the consequences of larger $\delta$ at the end of Sec.~\ref{sec:polarization control}.

Inside the ferroelectric, the electric field
\begin{equation}
    \frac{U^{es}(0)-U^{es}(d)}{d}=\frac{P-\sigma_s}{\varepsilon_f\varepsilon_0}
    \label{eq:electric_field_FE}
\end{equation}
is constant and $\varepsilon_f$ is the ferroelectric dielectric constant. $P>0$ indicates polarization pointing toward $I_2$, and the electric fields in the insulating spacer layers follow:
\begin{equation}
    \begin{aligned}
        \frac{U^{es}(0)-U^{es}(-l_1)}{l_1}=\frac{\sigma_s}{\varepsilon_1\varepsilon_0}, \\
        \frac{U^{es}(l_2+d)-U^{es}(d)}{l_2}=\frac{\sigma_s}{\varepsilon_2\varepsilon_0}, \label{eq:electric_field_I}
    \end{aligned}
\end{equation}
with $\varepsilon_{1,2}$ the dielectric constants of $I_{1,2}$.

Solving Eqs.~(\ref{eq:Thomas_Fermi_screening})-(\ref{eq:electric_field_I}) yields
\begin{align}
    \sigma_s &= \beta P, \label{eq:screening_charge}\\
    U^{es}(0) &= \frac{1}{\varepsilon_0}\!\left(\delta+\frac{l_1}{\varepsilon_1}\right)\!\beta P, \label{eq:electrostatic_potential_0}\\
    U^{es}(d) &= \frac{1}{\varepsilon_0}\!\left(\delta+\frac{l_2}{\varepsilon_2}\right)\!\beta P, \label{eq:electrostatic_potential_d}
\end{align}
with the form factor
\begin{equation}
    \beta = \frac{d/\varepsilon_f}{d/\varepsilon_f + l_1/\varepsilon_1 + l_2/\varepsilon_2 + 2\delta}.
    \label{eq:beta}
\end{equation}

Reversing the polarization changes the sign of $P$ and thus of $U^{es}(x)$ (Fig.~\ref{fig:system}). For thin barriers with large dielectric constant such that $l_1/\varepsilon_1,\, l_2/\varepsilon_2 \ll \delta$, Eqs.~(\ref{eq:screening_charge})-(\ref{eq:electrostatic_potential_d}) simplify to
\begin{align}
    \sigma_s &= \frac{d/\varepsilon_f}{d/\varepsilon_f+2\delta} P, \label{eq:screening_charge_simplified}\\
    U^{es}(0) &= -U^{es}(d)= \frac{d/\varepsilon_f}{d/\varepsilon_f+2\delta}\frac{\delta}{\varepsilon_0} P \equiv \Delta_p. \label{eq:electrostatic_potential_simplified}
\end{align}

In this limit $U^{es}$ is constant across $I_{1,2}$. In the following, for simplicity we focus on this experimentally relevant regime.

\section{Polarization-Controlled critical current and Tunable On-off Efficiency}
\label{sec:polarization control}

The supercurrent $I_s = I_c \sin \phi$ depends on the phase difference $\phi$ across the Josephson junction. At zero temperature, the critical current density  $J_c=I_c/A$ of a short tunnel junction with area A reads \cite{tinkham2004,ambegaokar1963}
\begin{equation}
    J_c = \frac{\pi \Delta_0}{2e} \frac{G_N}{A},
    \label{eq:critical_current}
\end{equation}
where $\Delta_0$ is the superconducting gap and $G_N=(2e^2/h)\sum_n T_n$ is the normal-state conductance expressed as a sum over spin-degenerate transmission eigenvalues $T_n$. Eq.~\eqref{eq:critical_current} holds for junctions with a weak link shorter than the superconducting coherence length, which is the case for our nanometer-thin barriers.

The Landauer formula for the conductance reads in the free-electron approximation:
\begin{equation}
    \frac{G_N}{A} = \frac{2 e^2}{h} \int_{0}^{k_F} \frac{k_{||} \, dk_{||}}{2\pi}\, T(E,k_{||}),
    \label{eq:Landauer}
\end{equation}
where $k_{||}=\sqrt{k_y^2+k_z^2}$ and $k_F$ is the Fermi wave vector. Numerical calculations of the transmission $T(E,k_{||})$ through the composite barrier for high density metals suffer from rapid oscillations that average out to a large extent in the integral. These oscillation are suppressed in the Wentzel-Kramers-Brillouin (WKB)  approximation \cite{simmons1963}:
\begin{equation}
    \begin{split}
        &T(E,k_{||}) \approx \\
        &\exp\left[-2 \int_{-l_1}^{d+l_2} \frac{\sqrt{2m\left(E_F + U - (E - \frac{\hbar^2 k_{||}^2}{2m})\right)}}{\hbar} dx \right] \label{eq:WKB}
    \end{split}
\end{equation}
that holds for sufficiently large potential barriers for which the square root is real.

The total potential $U(x)$ depends on the the electric polarization $P$, and so does  $J_c$. Two figures of merit are the (dimensionless) on-off efficiency
\begin{equation}
    \eta(P) = \frac{|J_c(+P)-J_c(-P)|}{J_c(+P)+J_c(-P)} \le 1,
\end{equation}
that measures the relative change of $J_c$ upon polarization reversal, and the polarization-averaged critical current density,
\begin{equation}
    \bar{J}_c(P)=\frac{J_c(+P)+J_c(-P)}{2}.
\end{equation}

Unless stated otherwise we use the following material parameters: superconducting gap $\Delta_0 = 2$ \si{meV} (reasonable for elemental s-wave superconductors such as Nb and Pb \cite{richards1960}); Fermi energy $E_F = 5$ \si{eV} (Nb \cite{Prozorov2022}); transmission energy at $E=E_F$; Thomas-Fermi screening length in high density metals $\delta=0.1$ \si{nm} \cite{kittel2018}; ferroelectric thickness $d=2$ \si{nm}; and ferroelectric intrinsic barrier $U_f^b=0.1$ \si{eV} \cite{rodriguezcontreras2003}. The ferroelectric dielectric constant is considered highly material and temperature dependent, thus we take a moderate value of $\varepsilon_f = 100$ \cite{zhuravlev2009} representative for ferroelectric perovskites and comment on the consequences of smaller and larger values at the end of this section. 

\begin{figure}[ht]
    \centering
    \includegraphics[width=0.95\columnwidth]{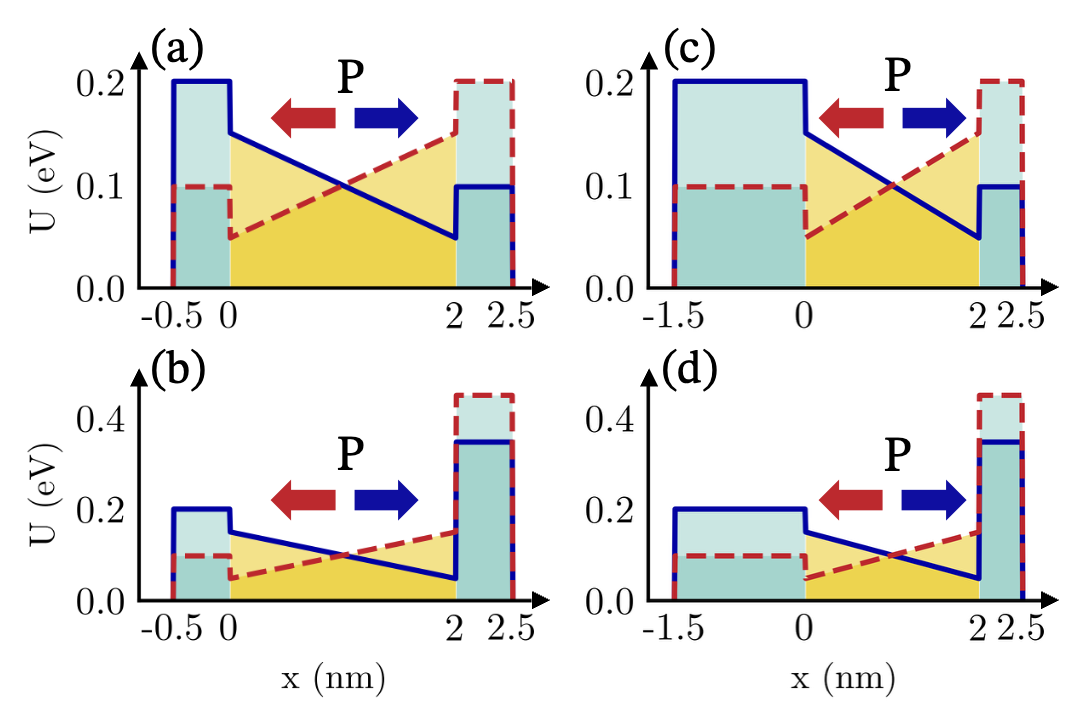}
    \caption{\textbf{Potential profiles.} Ferroelectric (yellow) and dielectric (blue) layers. Blue (solid) and red (dashed) curves: total potentials for opposite polarization directions (arrows). (a) Symmetric junction: $l_1=l_2=0.5$ \si{nm}, $U_1^b=U_2^b=0.15$ \si{eV}. (b) Barrier-asymmetric: $l_1=l_2=0.5$ \si{nm}, $U_1^b=0.15$ \si{eV}, $U_2^b=0.4$ \si{eV}. (c) Thickness-asymmetric: $U_1^b=U_2^b=0.15$ \si{eV}, $l_1=1.5$ \si{nm}, $l_2=0.5$ \si{nm}. (d) Strongly asymmetric: $l_1=1.5$ \si{nm}, $l_2=0.5$ \si{nm}, $U_1^b=0.15$ \si{eV}, $U_2^b=0.4$ \si{eV}.}
    \label{fig:potentials}
\end{figure}

\begin{figure}[ht]
    \centering
    \includegraphics[width=1\columnwidth]{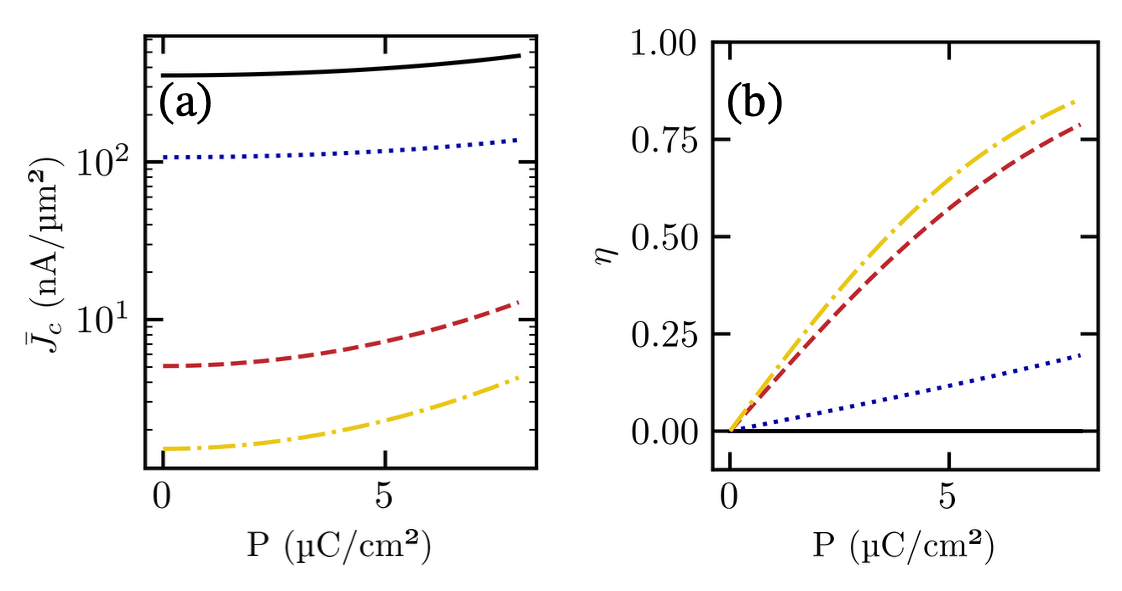}
    \caption{\textbf{Polarization dependence.} (a) Averaged critical current density $\bar{J}_c$ and (b) on-off efficiency $\eta$ vs polarization $P$. Solid black: symmetric junction of Fig.~\hyperref[fig:potentials]{2(a)}. Dotted blue: barrier asymmetry [Fig.~\hyperref[fig:potentials]{2(b)}]. Dashed red: thickness asymmetry [Fig.~\hyperref[fig:potentials]{2(c)}]. Dash-dot yellow: combined strong asymmetry [Fig.~\hyperref[fig:potentials]{2(d)}].}
    \label{fig:P_dependence}
\end{figure}

Fig.~\ref{fig:P_dependence} shows $\bar{J}_c(P)$ and $\eta(P)$ for the four representative geometries of Fig.~\ref{fig:potentials}. In the symmetric junction [Fig.~\hyperref[fig:potentials]{2(a)}] inversion symmetry guarantees $J_c(+P)=J_c(-P)$ and thus $\eta=0$. Introducing asymmetric potential barriers [Fig.~\hyperref[fig:potentials]{2(b)}] lowers $\bar{J}_c$ and monotonically raises $\eta$ with $|P|$. Different thicknesses but equal potential barriers [Fig.~\hyperref[fig:potentials]{2(c)}] lead to the same conclusion. Combining thickness and barrier asymmetries [Fig.~\hyperref[fig:potentials]{2(d)}] enhances the efficiency up to  $\eta \approx 0.9$ for $P = 8$ \si{\micro\coulomb/\centi\meter^{2}}, but at the cost of a small critical current density of a few \si{\nano\ampere/\micro\meter^{2}}.

\begin{figure}[ht]
    \centering
    \includegraphics[width=1\columnwidth]{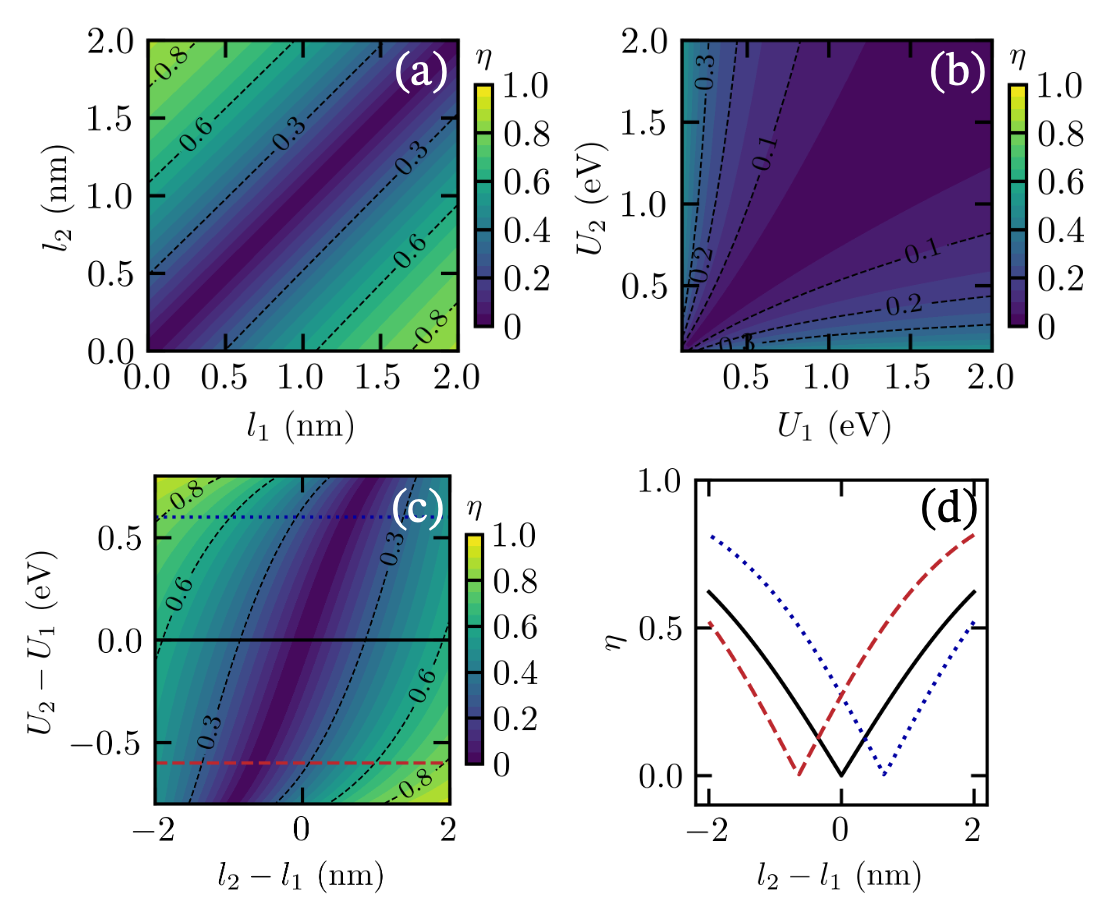}
    \caption{\textbf{Tunable on-off efficiency at fixed }$|P|=5$ \si{\micro\coulomb/\centi\meter^{2}}. (a) $\eta$ vs $l_1,l_2$ for $U_1^b=U_2^b=0.15$ \si{eV}. (b) $\eta$ vs $U_1^b,U_2^b$ for $l_1=l_2=1.0$ \si{nm}. (c) $\eta$ vs thickness difference $l_2-l_1$ and barrier difference $U_2^b-U_1^b$ at fixed $l_1+l_2=2$ \si{nm} and $U_1^b+U_2^b=1$ \si{eV}. (d) Line cuts of (c): $\eta$ vs $l_2-l_1$ for $U_2^b-U_1^b=0$ (solid black), $+0.6$ \si{eV} (dotted blue), and $-0.6$ \si{eV} (dashed red).}
    \label{fig:efficiency}
\end{figure}

We assess the tunability for fixed $|P|=5$ \si{\micro\coulomb/\centi\meter^{2}} and different structural parameters. Fig.~\hyperref[fig:efficiency]{4(a,b)} shows that the efficiency improves with both thickness and barrier asymmetry. Varying both simultaneously [Fig.~\hyperref[fig:efficiency]{4(c)}] can cancel the modulation efficiency for some parameter combinations: the line cuts in Fig.~\hyperref[fig:efficiency]{4(d)} emphasize that for \(l_1 \ne l_2\) the efficiency goes to zero. On the other hand, $\eta$ is maximized when the thinner (thicker) insulating layer has a larger (smaller) potential barrier, respectively.

\begin{figure}[ht]
    \centering
    \includegraphics[width=1\columnwidth]{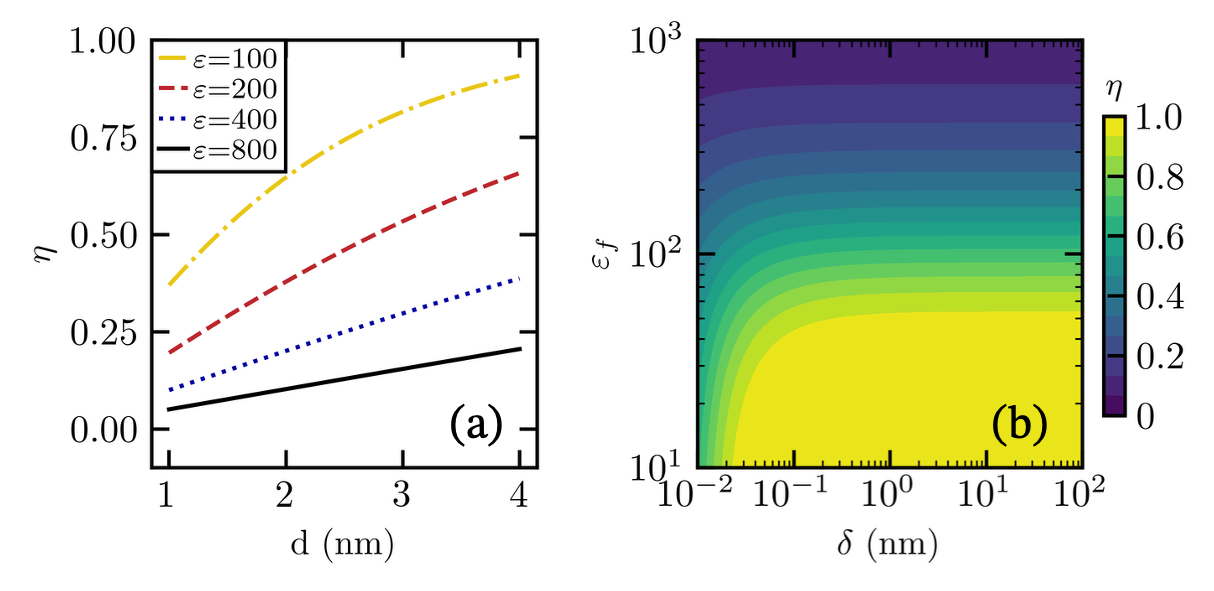}
    \caption{\textbf{Dependence on ferroelectric thickness, dielectric constant and screening length.} (a) $\eta$ vs. ferroelectric thickness $d$ for a strongly asymmetric junction in Fig.~\hyperref[fig:potentials]{2(d)} and dielectric constants $\varepsilon_f=800$ (solid black), $400$ (dotted blue), $200$ (dashed red), $100$ (dash-dot yellow). The screening length $\delta=0.1$ \si{nm}. (b) $\eta$ vs. screening length $\delta$ and $\varepsilon_f$ for $d=2$ \si{nm}.}
    \label{fig:eta-material}
\end{figure}

Fig.~\hyperref[fig:eta-material]{5(a)} illustrates that the efficiency increases with ferroelectric thickness $d$ and decreases with dielectric constant $\varepsilon_f$, consistent with an electrostatic potential change that scales like $\Delta_p \propto 1/(\varepsilon_f/d + \text{const.})$ [Eq.~(\ref{eq:electrostatic_potential_simplified})]. Thicker ferroelectric layers enhance the internal potential drop and hence the contrast between $\pm P$, but large total barrier thickness also suppresses the modulus of the critical current. A smaller dielectric constant increases the efficiency, but also implies a smaller $|J_c|$. Figure~\hyperref[fig:eta-material]{5(b)} examines the role of the screening length $\delta$. By rewriting Eq.~(\ref{eq:electrostatic_potential_simplified}) as

\begin{equation}
    \Delta_p = \frac{1/\varepsilon_0}{2\varepsilon_f/d + 1/\delta} P,
\end{equation}
we see that for large $\varepsilon_f$, as in many ferroelectric perovskites, the ferroelectric contribution dominates the electrostatic potential and $\eta$ is largely insensitive to $\delta$. For materials with smaller $\varepsilon_f$ (for example \ce{CuInP2S6} \cite{belianinov2015,zhou2020a,zhou2020b} or \ce{HfO2} \cite{park2015,park2018,schroeder2022}), the ferroelectric control of the potential is weaker and $\eta$ tends to benefit from a larger screening length [Eq.~(\ref{eq:electrostatic_potential_simplified})], although this dependence is modest in realistic regimes ($\delta \gtrsim 0.1$ \si{nm}). A comprehensive optimization of $\eta$ and $J_c$ across realistic layer structures, susceptibilities, and material polarizations is beyond the scope of this work.

\section{Linear Approximation of critical current}
\label{sec:analytical}

Here we study a small-parameter expansion of the polarization-dependent critical current starting from Eq.~(\ref{eq:Landauer}) and Eq.~(\ref{eq:WKB}).

The integral in Eq.~(\ref{eq:WKB}) can be separated into contributions from the three layers:
\begin{equation}
    T(k_{||}) \approx e^{-2(\gamma_1 + \gamma_2 + \gamma_{f})},
\end{equation}
where
\begin{align}
    \gamma_1 &= l_1\sqrt{\frac{2mU_1}{\hbar^2}+k_{||}^2}, \\
    \gamma_2 &= l_2\sqrt{\frac{2mU_2}{\hbar^2}+k_{||}^2}, \\
    \begin{split}
        \gamma_{f} &= \frac{2}{3}\frac{d\hbar^2}{2m\left [U_f(d)-U_f(0)\right ]} \\ & \quad \times \left \{ \left [\frac{2mU_f(d)}{\hbar^2} + k_{||}^2\right ]^{\frac{3}{2}} - \left [\frac{2mU_f(0)}{\hbar^2} + k_{||}^2\right ]^{\frac{3}{2}} \right \},
    \end{split}
\end{align}
are the WKB exponents for the two insulating layers and the ferroelectric layer, respectively. $U_1 = U^b_1 + \Delta_p$ and $U_2 = U^b_2 - \Delta_p$ are the total potential barriers of the insulating layers, while $U_f(0) = U^b_f + \Delta_p$ and $U_f(d) = U^b_f - \Delta_p$ are those at their interfaces to the ferroelectric and $\Delta_p$ is the polarization-induced electrostatic offset in Eq.~(\ref{eq:electrostatic_potential_simplified}).

For sufficiently small small $k_{||}$ or high barriers $k_{||}^2 \ll 2mU/\hbar^2$, we may expand the square roots to obtain
\begin{align}
    \gamma_1 &\approx \frac{l_1\sqrt{2mU_1}}{\hbar} + \frac{1}{2}\frac{l_1\hbar}{\sqrt{2mU_1}}k_{||}^2 , \\
    \gamma_2 &\approx \frac{l_2\sqrt{2mU_2}}{\hbar} + \frac{1}{2}\frac{l_2\hbar}{\sqrt{2mU_2}}k_{||}^2 , \\
    \begin{split}
        \gamma_f &\approx \frac{2}{3} \frac{d\sqrt{2m}\left [ U_f(0)^{3/2} - U_f(d)^{3/2}\right ]}{\hbar\left [ U_f(0)-U_f(d) \right ]} \\ 
        & \quad + \frac{d\hbar \left [ \sqrt{U_f(0)}-\sqrt{U_f(d)}\right ]}{\sqrt{2m} \left [ U_f(0) - U_f(d) \right ]} k_{||}^2 .
    \end{split}
\end{align}

When additionally $\Delta_p / U^b \ll 1$, the transmission takes the form
\begin{equation}
    T(k_{||}, \Delta_p) = e^{-(A_p+B_pk_{||}^2)},
\end{equation}
where $A_p$ and $B_p$ are polarization-dependent coefficients:
\begin{align}
    \begin{split}
        A_p &= 2 \frac{\sqrt{2m}}{\hbar} \left[ l_1\sqrt{U_1^b} + l_2\sqrt{U_2^b} + d\sqrt{U_f^b} \right. \\
        &\left. \quad - \frac{1}{2} \left ( \frac{l_2}{\sqrt{U_2^b}} - \frac{l_1}{\sqrt{U_1^b}} \right )\Delta_p \right],
    \end{split}\\
    B_p &= \frac{\hbar}{\sqrt{2m}} \left\{ \frac{l_1}{\sqrt{U_1^b}} + \frac{l_2}{\sqrt{U_2^b}} + \frac{d}{\sqrt{U_f^b}} \right. \\
    &\left. \quad +\frac{1}{2}\left[ \frac{l_2}{(U_2^b)^{\frac{3}{2}}} - \frac{l_1}{(U_1^b)^{\frac{3}{2}}} \right] \Delta_p \right\}.
\end{align}

\begin{figure}[ht]
    \centering
    \includegraphics[width=0.75\columnwidth]{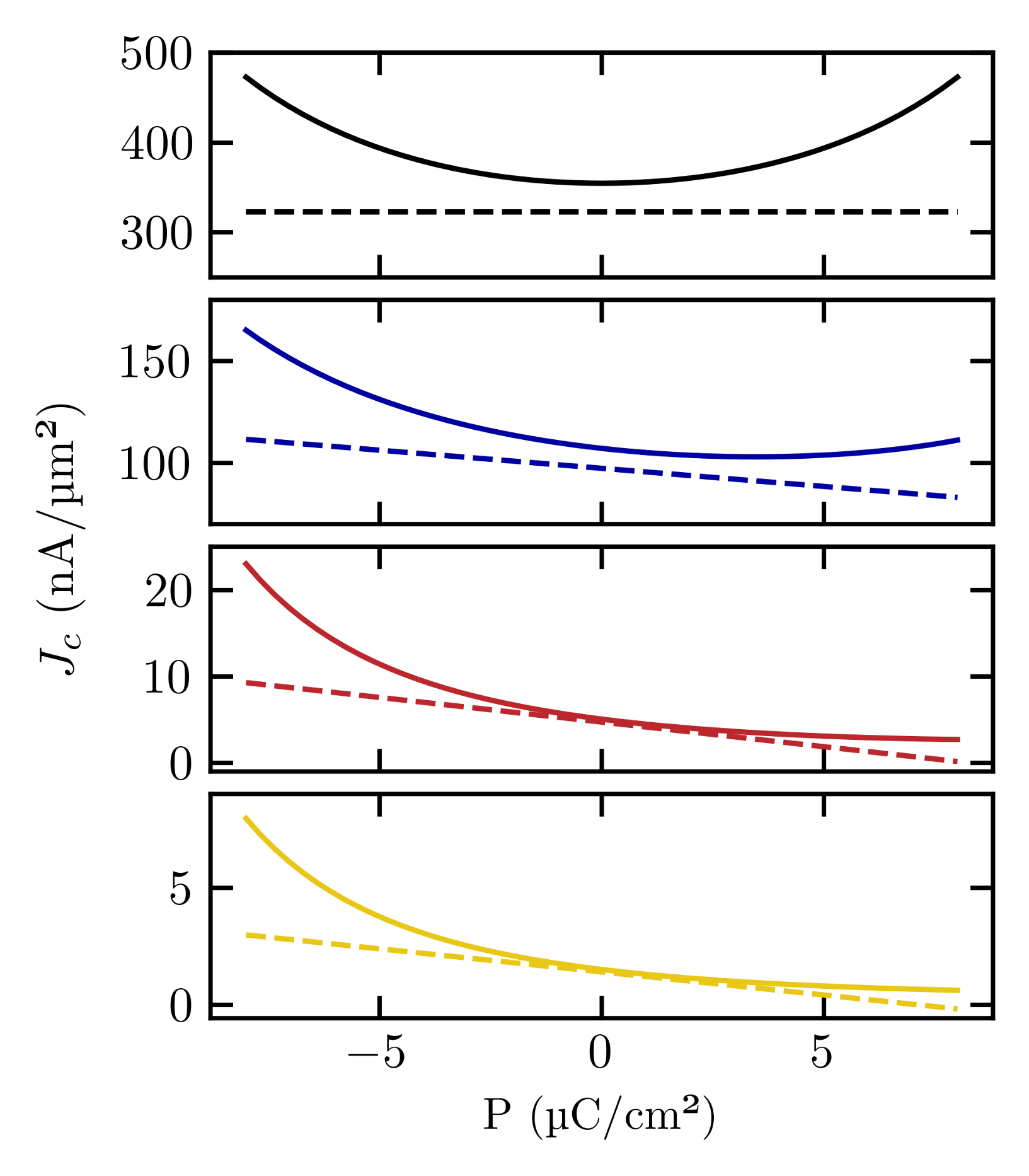}
    \caption{\textbf{Linear approximation of the polarization dependence of the critical current density.} Critical current density $J_c$ vs polarization $P$. (a)-(d) correspond to the junctions in Fig.~\hyperref[fig:potentials]{2(a)}-\hyperref[fig:potentials]{(d)}. Solid: numerical results; dashed: analytical approximation.}
    \label{fig:linear_approximation}
\end{figure}

In this limit the Landauer integral Eq.~(\ref{eq:Landauer}) can be evaluated analytically when $k_F$ is much larger than the scale over which the Gaussian-like integrand $k_{||} e^{-(A_p+B_p k_{||}^2)}$ vanishes. We introduce a cutoff $k_c \equiv 1/\sqrt{B_p}$, for which the small $k_{||}$ approximation holds. The critical current density then simplifies to
\begin{equation}
    \begin{split}
        J_c &= \frac{e\Delta_0}{2h} \frac{e^{-A_p}}{2B_p} \left[ 1-e^{-B_pk_c^2} \right] \\
        & \approx \frac{e\Delta_0}{2h} \frac{e^{-A_p}}{2B_p}. \label{eq:critical_current_approx}
    \end{split}
\end{equation}

To first order in $\Delta_p \sim P$ we getfvv                                                                                                                                                                                                                                                                                                                   

\begin{equation}
    J_c (P) = \frac{e\Delta_0}{4h} \kappa \left( 1+ \theta P \right),
    \label{eq:critical_current_linear}
\end{equation}
and 
\begin{equation}
    \eta(P) = \lvert \theta P \rvert,
    \label{eq:efficiency_linear}
\end{equation}
where $\kappa$ and $\theta$ are device-dependent coefficients:
\begin{align}
    \kappa &= \frac{\exp\left\{-2\frac{\sqrt{2m}}{\hbar}\left( l_1\sqrt{U_1^b} + l_2\sqrt{U_2^b} + d\sqrt{U_f^b} \right)\right\}}{\frac{\hbar}{\sqrt{2m}}\left( l_1/\sqrt{U_1^b} + l_2/\sqrt{U_2^b} + d/\sqrt{U_f^b} \right)}, \\
    \begin{split}
        \theta &= \frac{d\delta/\epsilon_0}{d+2\delta\epsilon_f} \left[\frac{l_1/(U_1^b)^{3/2}-l_2/(U_2^b)^{3/2}}{l_1/\sqrt{U_1^b}+l_2/\sqrt{U_2^b}+d/\sqrt{U_f^b}} \right. \\
        &\left. \quad -\frac{\sqrt{2m}}{\hbar} \left( \frac{l_1}{U_1^b} -\frac{l_2}{U_2^b} \right)\right].
    \end{split}
\end{align}

Fig.~\ref{fig:linear_approximation} compares the numerical and analytical results of the critical current density. The linear approximation captures both magnitude and trend for small $|P|$. The slight offset at $P=0$ arises from the small-$k_{||}$ assumption in the WKB integral and the approximation taken in the second step of Eq.~(\ref{eq:critical_current_approx}). The compact formulas in Eqs.~(\ref{eq:critical_current_linear}) and (\ref{eq:efficiency_linear}) thus provide rapid estimates of polarization control in ferroelectric Josephson junctions.

\section{Conclusions}
\label{sec:conclusion}

We analyzed a composite S-I\(_1\)-FE-I\(_2\)-S Josephson junction. When breaking inversion symmetry by different thicknesses and/or heights of the potential barriers that separate the ferroelectric from the superconductor, a ferroelectric polarization reversal strongly modulates the critical current. In a WKB approximation we compute an electrically controlled on-off efficiency up to $\eta \simeq 0.9$ for physically realistic parameters. The efficiency $\eta$ can be maximized by thicknesses and potential barrier heights that are larger and lower on one side than on the other, in thicker ferroelectric films with large polarization, but small dielectric constant. We derive a compact formula in the limit of small \(P\) for quick estimates and interpreting numerical results. We conclude that ferroelectric Josephson junctions are interesting candidates for electrically programmable superconducting current switches in cryogenic memories and logic circuits.

\section{Acknowledgments}
The authors acknowledge the support from the Kavli Foundation under the Kavli Institute Innovation Award (KIIA).

\bibliography{main}
\end{document}